\documentclass[twocolumn,showpacs,preprintnumbers,amsmath,amssymb]{revtex4}
\usepackage{amsmath,amsfonts,latexsym,amssymb,graphics,epsfig,subfigure,color,makeidx}
\usepackage{xcolor}
\usepackage[utf8]{inputenc}
\usepackage{graphicx,hyperref}
\usepackage{bm}
\usepackage{color}

\begin{document}
	
	\title{Gravitational radiation pulses from Extreme-Mass-Ratio-Inspiral system with a  supermassive boson star}
	
	\author{Yu-Peng Zhang$^{1,\,2}$\footnote{zyp@lzu.edu.cn},
		Yan-Bo Zeng$^{1,\,2}$\footnote{zengyb19@lzu.edu.cn},
		Yong-Qiang Wang$^{1,\,2}$\footnote{yqwang@lzu.edu.cn},
		Shao-Wen Wei$^{1,\,2}$\footnote{weishw@lzu.edu.cn},
		Pau Amaro Seoane$^{3,\,4,\,5,\,6}$\footnote{amaro@upv.es},
		Yu-Xiao Liu$^{1,\,2}$\footnote{liuyx@lzu.edu.cn, corresponding author}
	}
	\affiliation{$^{1}$Lanzhou Center for Theoretical Physics, Key Laboratory of Theoretical Physics of Gansu Province, School of Physical Science and Technology, Lanzhou University, Lanzhou 730000, China\\
		$^{2}$Institute of Theoretical Physics \& Research Center of Gravitation, Lanzhou University, Lanzhou 730000, China\\
		$^{3}$Institute of Multidisciplinary Mathematics, Universitat Polit\`{e}cnica de Val\`{e}ncia, Spain\\
		$^{4}$DESY Zeuthen, Germany\\
		$^{5}$Kavli Institute for Astronomy and Astrophysics at Peking University, Beijing 100871, China\\
		$^{6}$Institute of Applied Mathematics, Academy of Mathematics and Systems Science, CAS, Beijing 100190, China
	}
	
	\begin{abstract}
		
		Future space-borne gravitational-wave detectors will observe the
		gravitational waves in the milli-Hz. Extreme-mass-ratio inspirals {with
			central} supermassive black holes are {very important sources} that could
		provide the information of the vicinity of black holes. The event horizon separates
		the {inner region} of a black hole and there is nothing that can escape from this
		region. When the central supermassive compact object is a regular and
		horizonless rotating boson star, a small body could pass through the center and
		{follow novel types of orbits}. {These} will generate the
		gravitational waves that can not be obtained in the scenario corresponding to an
		extreme-mass-ratio inspiral with a central supermassive black hole. {This can
			be} used to examine whether a supermassive rotating boson star is present
		at the centers of galaxies. In this work, we consider an
		extreme-mass-ratio inspiral system described by a small compact object
		inspiralling into a central supermassive rotating boson star. Integrating four
		types of special equatorial geodesics and using the numerical kludge method
		with quadrupole approximation, we obtain the corresponding gravitational
		waveforms and find that there are high-frequency gravitational radiation pulses
		in such system. The frequencies of {the} gravitational radiation pulses could
		be in the magnitude of $10^{-1}$Hz and the whole {gravitational wave} parts are
		in the milli-Hz. These novel gravitational waveforms are quite different from
		that produced by the extreme-mass-ratio inspiral system with a supermassive
		central black hole. By assuming the masses of {the} central supermassive
		rotating boson star and small compact object to be $10^6 M_\odot$ and $10
		M_\odot$ and assuming a distance of $1\text{Gpc}$, we show that the gravitational radiation pulses
		could be detected by the space-borne gravitational-wave
		detectors. Our results will provide a possible evidence to distinguish the
		astrophysical compact objects in the galactic centers.
		
	\end{abstract}

	\maketitle
	\section{Introduction}
	In 2015, LIGO successfully detected the gravitational waves (GWs) generated by the merger of binary black holes~\cite{Abbott2016a}, which declares that the era of {gravitational wave astronomy} is coming. {Besides LIGO} and VIRGO, there are several independent space-borne detectors on their way, such as the Laser
	Interferometer Space Antenna~\cite{lisapaper,lisawebsite}, Taiji~\cite{taiji}, Tianqin~\cite{Luo2016}, and DECIGO~\cite{Shuichi2017}. They might be launched into the space in the next ten years and the detectable frequency of GWs will be extended into the milli-Hz. Extreme-mass-ratio-inspirals (EMRIs)~\cite{Amaro-Seoane2018b} are the one type of most promise sources with the frequency in milli-Hz. They are described by the late capture orbits of stellar-mass compact objects inspiralling into central supermassive black holes in galactic centers. The corresponding GWs radiated from such systems are closely related to the accurate information in the vicinity of the central {objects}. {It is well known that} the existence of event horizon for a black hole is one of the most intriguing predictions of general relativity, which means there is nothing can escape from the inside of the event horizon. Thus, it can be considered as the merger of an EMRI system when the small star plunges into the event horizon of the central supermassive black hole. The final stage of the GWs will have gravitational radiation  pulses~\cite{Davis:1972ud} when the stellar-mass compact object plunges into the central supermassive black hole.
	
	Boson stars as the black hole mimickers had been proposed in the
	1970s~\cite{Feinblum1968,Kaup1968,Ruffini1969}, they are formed by the
	gravitationally bound Bose-Einstein condensates for light scalars with long de
	Broglie wavelengths. Boson stars could have
	spherical~\cite{Seidel1994,Giovanni2018} or rotating~\cite{Sanchis2019}
	configurations. Combining the observational data and the possible stellar
	distribution in the background of a central boson star, the possibility that
	whether a supermassive boson star can be in the center of a galaxy was
		discussed in Ref.~\cite{diego2000}. Later, the work of \cite{Amaro-SeoaneBarrancoBernalRezzolla10} exploits the orbits of the S-stars
	at the Galactic Centre to impose constraints to a hypothetical supermassive boson star. What are the possible observations that
	could be seen with a central supermassive boson star is still
	investigated~\cite{Yuan:2004sv}. By comparing the power spectrum of a simple
	accretion disk model between a spherical boson star and a Schwarzchild black
	hole, the result of Ref.~\cite{Rueda2009} {shows} that it is possible to find a
	boson star mimics the power spectrum of the disk of a black hole with the same
	mass. The shadow of a boson star still provides the possibility that a
	spherical Proca star could mimic a Schwarzchild black
	hole~\cite{Herdeiro2021lwl}. Therefore, it is necessary to study what is the
	possible GWs from the EMRIs with central {supermassive} rotating boson stars.
	
	As the space-borne gravitational-wave detection program continues to advance, the detection of the GWs from EMRIs is becoming possible. Boson star is a regular and horizonless strong gravity system, which means it is particular well motivated for considering a supermassive boson star located in the center of {a galaxy}. The horizonless property means that a small star could plunge into and pass through the center of a boson star. That is to say, the GWs generated by such system in the final stage will be quite different from the EMRIs with central supermassive black holes. Reference~\cite{Kesden2005} has proven that the novel GWs will exist in the central supermassive spherical boson star. Note that, for a rotating boson star, the novel geodesics that are different from a Kerr black hole have been obtained~\cite{philippe2014,Grould12017,Collodel2018,Yuzhang2021}. With such novel geodesic orbits, the corresponding GWs will be really critical to study the properties of boson stars. In this work, we will use these kind of orbits to obtain the corresponding possible novel GWs.
	
	To describe the gravitational mass scale of various boson stars, a natural length {scale} (the reduced Compton wavelength)
	\begin{equation}
	\bar{\lambda}=\frac{\hbar}{\mu c}
	\end{equation}
	is necessary, where $\mu$ is the scalar field mass parameter, $c$ is the speed of light. The order of {the magnitude} of {the boson star mass} related with $\bar{\lambda}$ is~\cite{Palenzuela2012}
	\begin{equation}
	M_{\text{bos}} \sim \frac{c^2~\bar{\lambda}}{G}=\frac{m^2_p}{\mu},\label{mass-of-boson-star}
	\end{equation}
	where the Planck mass $m_p=\sqrt{\hbar c/G}\approx 2.18\times 10^{-8}~ \text{kg}$. For a boson star with ultra-light mass parameter, the mass scale could be comparable with a supermassive black hole~\cite{Palenzuela2012,philippe2014}.
	
	In this work, we consider the EMRI system described by a stellar-mass compact object inspiralling into a central supermassive rotating boson star and investigate the corresponding GWs with the novel orbits. If the supermassive compact object located in the center of a galaxy is a boson star, the EMRI system will be different and what are the possible observable GWs will be very important. We will use the numerical kludge (NK) method~\cite{Stanislav2007} to obtain the approximate GWs emitted by a small object inspiralling into a central supermassive boson star.
	
	\section{Equatorial geodesics and Gravitational waves} \label{sec:fundamentals}
	{In order to construct the} equilibrium configurations {of rotating} boson stars, {we take the following ansatz for the metric~\cite{Herdeiro2014,Herdeiro2015}}
	\begin{eqnarray}
	ds^2&=&g_{tt}dt^2+g_{rr}dr^2+g_{\theta\theta}d\theta^2+2g_{t\varphi}dt d\varphi+g_{\varphi\varphi}d\varphi^2\nonumber\\
	&=&- e^{2F_0}dt^2
	+ e^{2F_1}\left(dr^2+r^2d\theta^2\right)\nonumber\\
	&& +~e^{2F_2}r^2\sin^2\theta(d\varphi-Wdt)^2.
	\label{metric-sbs}
	\end{eqnarray}
	We consider the mini boson star~\cite{Schunck2003} with following action
	\begin{equation}
	S =  \int{d^4 x \sqrt{-g}
		\left[\frac{R}{16\pi G} - \nabla_\mu\Phi\nabla^\mu\Phi^* - \frac{\mu^2}{\hbar^2}\Phi\Phi^*\right]},
	\label{action}
	\end{equation}
	where $G$ is {Newton's gravitational} constant and $m$ is the mass parameter of the scalar field. We set $G$, {the speed of light $c$, and the Planck constant $\hbar$} to be unity ($G=c=\hbar=1$). For a stationary and axisymmetric rotating boson star, the ansatz for the complex scalar field $\Phi$ is
	\begin{equation}
	\Phi=\phi(r,\theta)\exp\left[i(\omega t -k\varphi)\right],
	\label{scalarf}
	\end{equation}
	where $\omega$ is the frequency of the scalar field. The configuration of a equilibrium boson star could be obtained with the suitable boundary conditions, see the details in Ref.~\cite{Herdeiro2015}. Here, we only consider a rotating boson star with $k=1$ and $\omega=0.79 m $.
	
	A rotating boson star is stationary and axisymmetry, which means there are a timelike killing vector $\xi^\mu=(\partial_t)^\mu$ and a spacelike killing vector $\eta^\mu=(\partial_\varphi)^\mu$. These two independent {Killing vectors lead to} the {following two conserved quantities:}
	\begin{eqnarray}
	\bar{E} &=& -(\partial_t)^\mu u_{\mu}=-g_{tt}u^t-g_{t\varphi}u^\varphi,
	\label{energy} \\
	\bar{J} &=&(\partial_\varphi)^\mu u_{\mu}=g_{t\varphi}u^t+g_{\varphi\varphi}u^\varphi,
	\label{angularmomentum}
	\end{eqnarray}
	where $\bar{E}$ and $\bar{J}$ are the energy and angular momentum of the particle per {unit} mass. For simplicity, we only consider a test particle that moves in the equatorial orbits, {for which} the four-velocity is
	\begin{equation}
	u^{\mu}=\left(\frac{dt}{d\tau},\frac{dr}{d\tau},0,\frac{d\varphi}{d\tau}\right)=\left(u^t,u^r,0,u^\varphi\right),
	\end{equation}
	where $\tau$ is the proper time of the test particle. Solving Eqs.~\eqref{energy}-\eqref{angularmomentum} {and $u^\mu u_{\mu}=-1$, we have the following non-vanishing components of the velocity}
	\begin{eqnarray}
	u^t &=&\frac{\bar{E} g_{\varphi\varphi} + \bar{J} g_{t\varphi}}{g_{t\varphi}^2 - g_{tt}g_{\varphi\varphi}},\label{ut_eq} \\
	u^\varphi &=& \frac{\bar{E} g_{t\varphi} + \bar{J} g_{tt}}{g_{tt}g_{\varphi\varphi} - g_{t\varphi}^2},\label{uphi_eq}\\
	\left(u^r\right)^2 &=& -\frac{1+g_{\varphi\varphi}u^\varphi u^\varphi + 2 g_{t\varphi}u^t u^\phi + g_{tt}u^t u^t}{g_{rr}}.
	\label{velocityur}
	\end{eqnarray}

	Using the effective potential is an easy way to determine the radial motion of a test particle in a central field. {By decomposing the form of Eq. \eqref{velocityur} as}
	\begin{eqnarray}
	(u^r)^2 \propto \left(\bar{E}-V^{+}_{\text{eff}}\right)\left(\bar{E}-V^{-}_{\text{eff}}\right)
	\label{effectivepotentiala}
	\end{eqnarray}
	{we can obtain} the effective potentials $V^{+}_{\text{eff}}$ and $V^{-}_{\text{eff}}$ of a test particle in the background of a rotating boson star~\cite{Grould12017,Collodel2018,Yuzhang2021}. By checking the relation between the effective potentials and angular momentum $\bar{J}$, one can prove that a test particle could be in a circular orbit with arbitrary angular momentum and there is no innermost stable circular orbit for the test particle in a rotating boson star. This special property leads to that the velocity of {the test particle} in the rotating boson star satisfies
	\begin{equation}
	(u^r, u^\varphi)=(0, 0)
	\end{equation}
	{at some special points where} the angular momentum and energy satisfy the following condition~\cite{Collodel2018}
	\begin{equation}
	\bar{E}=-\frac{\bar{J} g_{tt}}{g_{t\varphi}}.
	\label{vphicondition}
	\end{equation}
	Combining the effective potentials \eqref{effectivepotentiala} and \eqref{vphicondition} and choosing suitable energy and angular momentum for the small body, we can obtain the novel orbits {that do not appear in the Kerr black hole case}.

	In this work, we consider four types of special orbits in a rotating boson star. {These novel orbits can be used to distinguish boson stars from black holes.} We give the periastron $r_p$, {apastron $r_a$, angular momentum} $\bar{J}$, and energy $\bar{E}$ of {these} orbits in Table~\ref{energy_angular_momentum_each_orbits}. With the help of the definition of the orbital eccentricity
	\begin{equation}
	e=\frac{r_a-r_p}{r_a+r_p},
	\end{equation}
	it is easy to prove that the orbital eccentricity $e$ for a test particle in a rotating boson star could be $1$, this result is unacceptable for a black hole.
	
\begin{figure}[htbp]
		\begin{center}
			\includegraphics[width=1.0\linewidth]{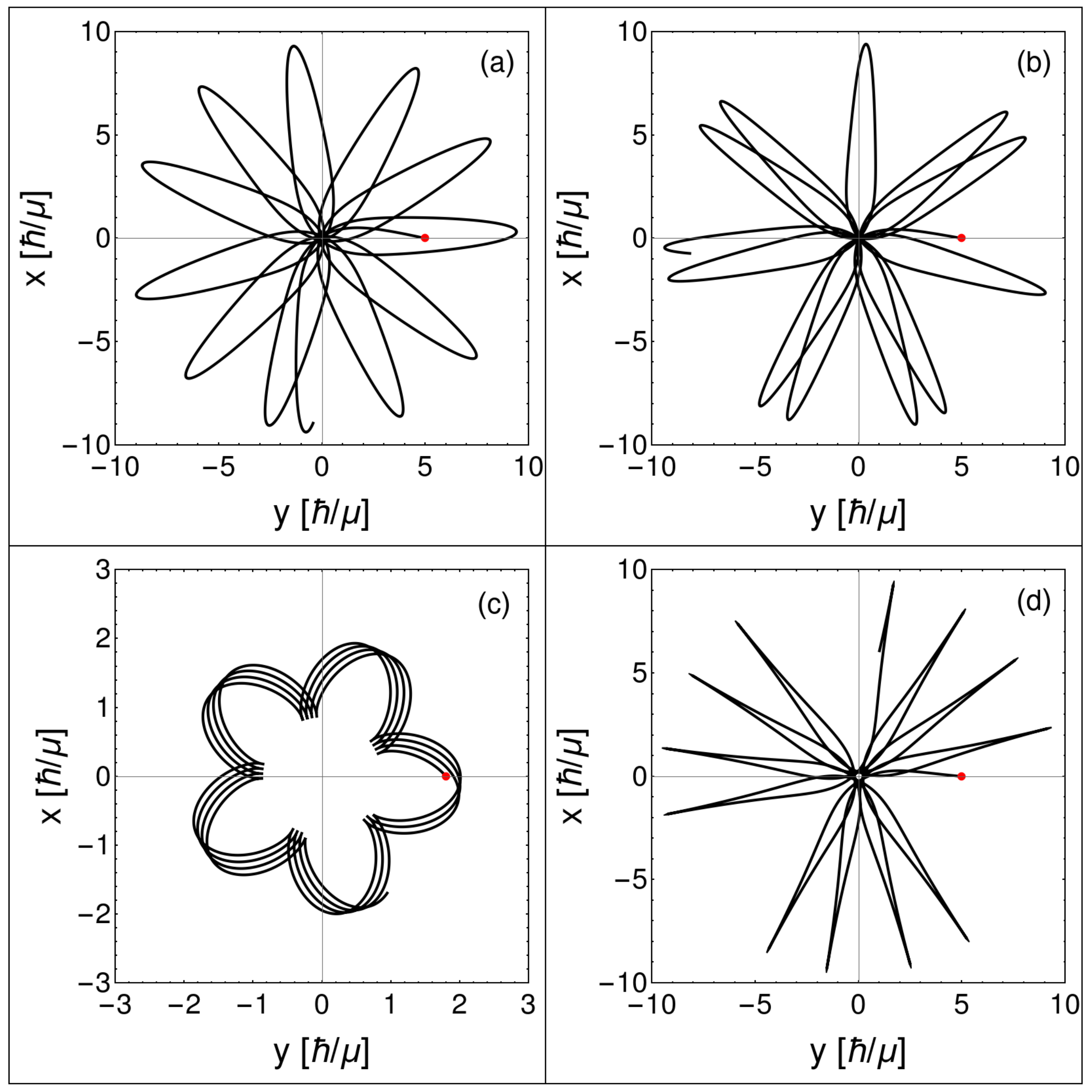} \hspace{5mm}
		\end{center}
		\vspace{-5mm}
		\caption{Four special types of orbits of {a small body} in a rotating boson star.}
		\label{orbits-123}
	\end{figure}

	{Figure}~\ref{orbits-123} shows the corresponding orbits that we considered. For the orbit (a), the periastron $r_p$ is less than the corresponding effective Schwarzchild radius of {the boson star}. For the orbit (b), the test particle could pass through the center of the rotating boson star and the periastron $r_p=0$. For the orbit (c), the radial velocity and angular velocity will be zero when the test particle is located at the periastron. For the orbit (d), the radial velocity and angular velocity will be zero when the test particle is located at the apastron. The orbits (a), (b), and (d) have the same apastrons but different periastrons. These {four  of} special orbits will offer  possible GW signals to check whether {spuermassive boson stars exist}.
	
	\begin{table}[!htb]
		\begin{center}
			\caption{Values of the parameters for the four {types} of orbits.}
			\begin{tabular}{ c| c |c |c |c |c}
				\hline
				\hline
				~orbit~&~~$~\bar{J}~$~~&~~~~$r_a/M$~~~~&~~~~$r_p/M$~~~~&~~~~$\bar{E}$~~~~&~~~~$e$~~~~\\
				\hline
				a      &  0.15         &   9.41476   &  0.10650    &  0.86788     &  0.97735         \\
				b      &  0            &   9.41476   &  0          &  0.86746     &  1               \\
				c      & -0.15         &   2.07405   &  0.92260    &  0.61943     &  0.38426         \\
				d      & -0.30         &   9.41476   &  0.20175    &  0.86713     &  0.95777         \\
				\hline
				\hline
			\end{tabular}
			\label{energy_angular_momentum_each_orbits}
		\end{center}
	\end{table}

	Next, we compute the corresponding GWs of the stellar-mass compact object inspiralling into a supermassive rotating boson star with the orbits given in Fig.~\ref{orbits-123}. In general, the GWs from an EMRI can be determined accurately by using the perturbation theory~\cite{Teukolsky1973,Hughes:2001jr}. {For a rotating boson star}, the background solution is numerical and it is really hard to obtain the gravitational perturbation equations. {Therefore, we} use the numerical kludge (NK) method~\cite{Stanislav2007} to generate the corresponding GWs for the EMRI system with a central supermassive rotating boson star.
	
	Using the well-known solutions of the linearized gravitational perturbation and combining the quadrupole formula in terms of the geodesic trajectory, we could get the approximate GWs. By integrating the four-velocity, we can obtain the corresponding trajectories. Once the trajectories of the small body have been obtained, one can construct the effective trajectories by projecting the Boyer-Lindquist coordinates into a pseudo-flat space as follows
	\begin{equation}
	x=r\sin\theta\cos\varphi,~y=r\sin\theta\sin\varphi,~ z=r\cos\theta.
	\label{cartesian-coordinate}
	\end{equation}
	Then we can obtain the GWs with the help of the geodesic trajectories and the ``quadrupole approximation"
	\begin{equation}
	\bar{h}^{ij}=\frac{2G}{d}\ddot{I}^{jk}|_{t^\prime=r-d},
	\label{eq-gws}
	\end{equation}
	where the parameter $d$ is the distance from the source to the observation point. The definition of {the} quadrupole $I^{jk}$ is
	\begin{equation}
	I^{jk}(t^\prime)=\int{x^{\prime j}x^{\prime k}T^{00}(t^{\prime},x^{\prime})d^3x^{\prime}},
	\end{equation}
	where the coordinate $x^{\prime j}$ is the position of a small star along the corresponding geodesic trajectory described by \eqref{cartesian-coordinate}. We compute the GWs of the cross and plus polarizations by adopting the Transverse-Traceless projection result \eqref{eq-gws} based on Ref.~\cite{Kesden2005}.

	{The calculation of the GWs is done with the NK method. With such approximation, the corresponding amplitudes of the GWs will be inaccurate. Therefore, our calculation is  ``phenomenological" but the frequencies of the GWs are correct.} Reference~\cite{Kesden2005} showed that the kludge waveforms are extremely well at approximating the true gravitational waveform and they have overlaps with the Teukolsky waveforms of $95 \%$~\cite{Kesden2005}. That is to say, it is possible to use these {kludge waveforms} with achievable quality to provide some roles in the search of space-borne gravitational-wave data for EMRIs with central supermassive rotating boson stars.
	
\begin{widetext}
		
		\begin{figure}[htbp]
			\begin{center}
				\includegraphics[width=0.49\linewidth]{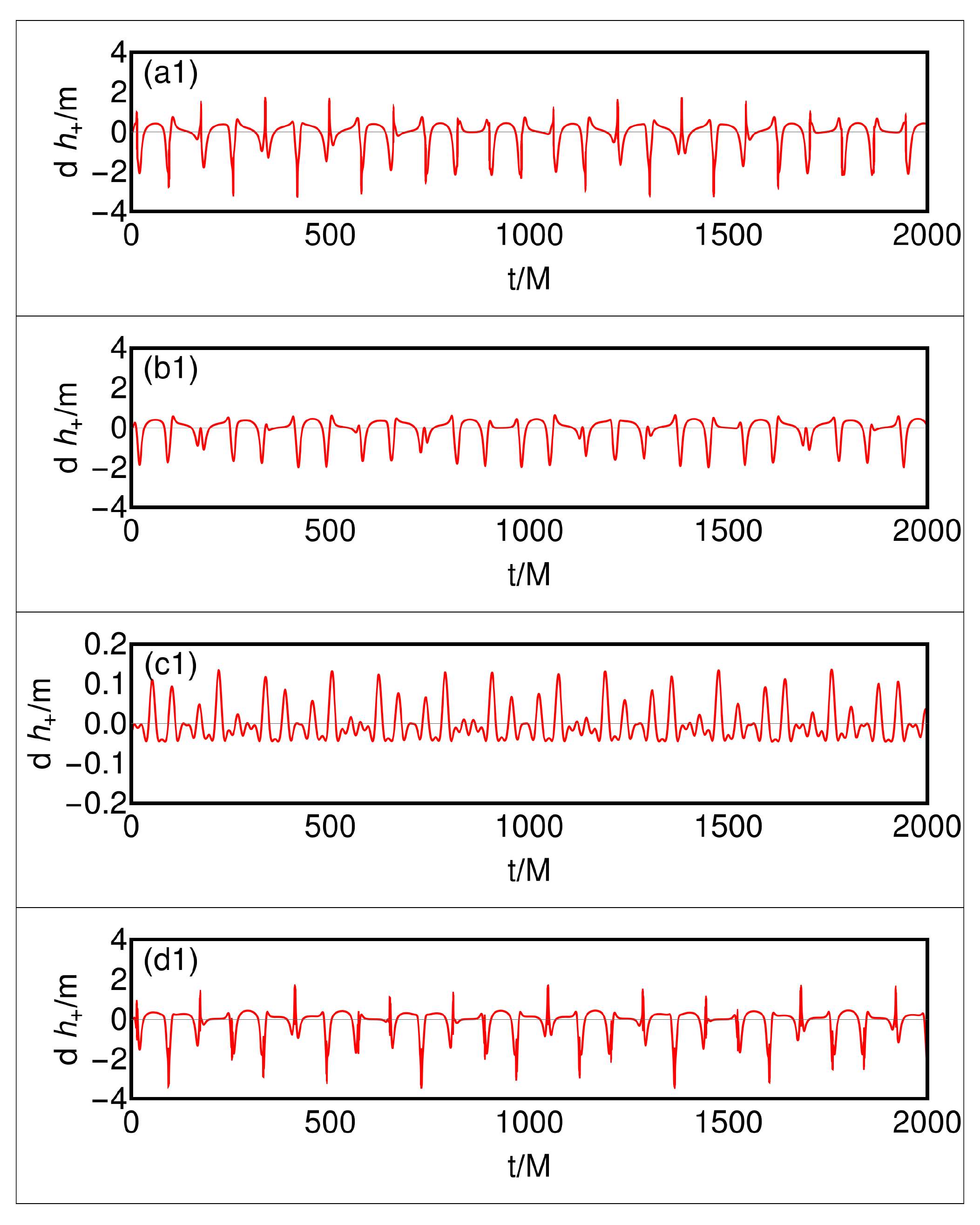}
				\includegraphics[width=0.49\linewidth]{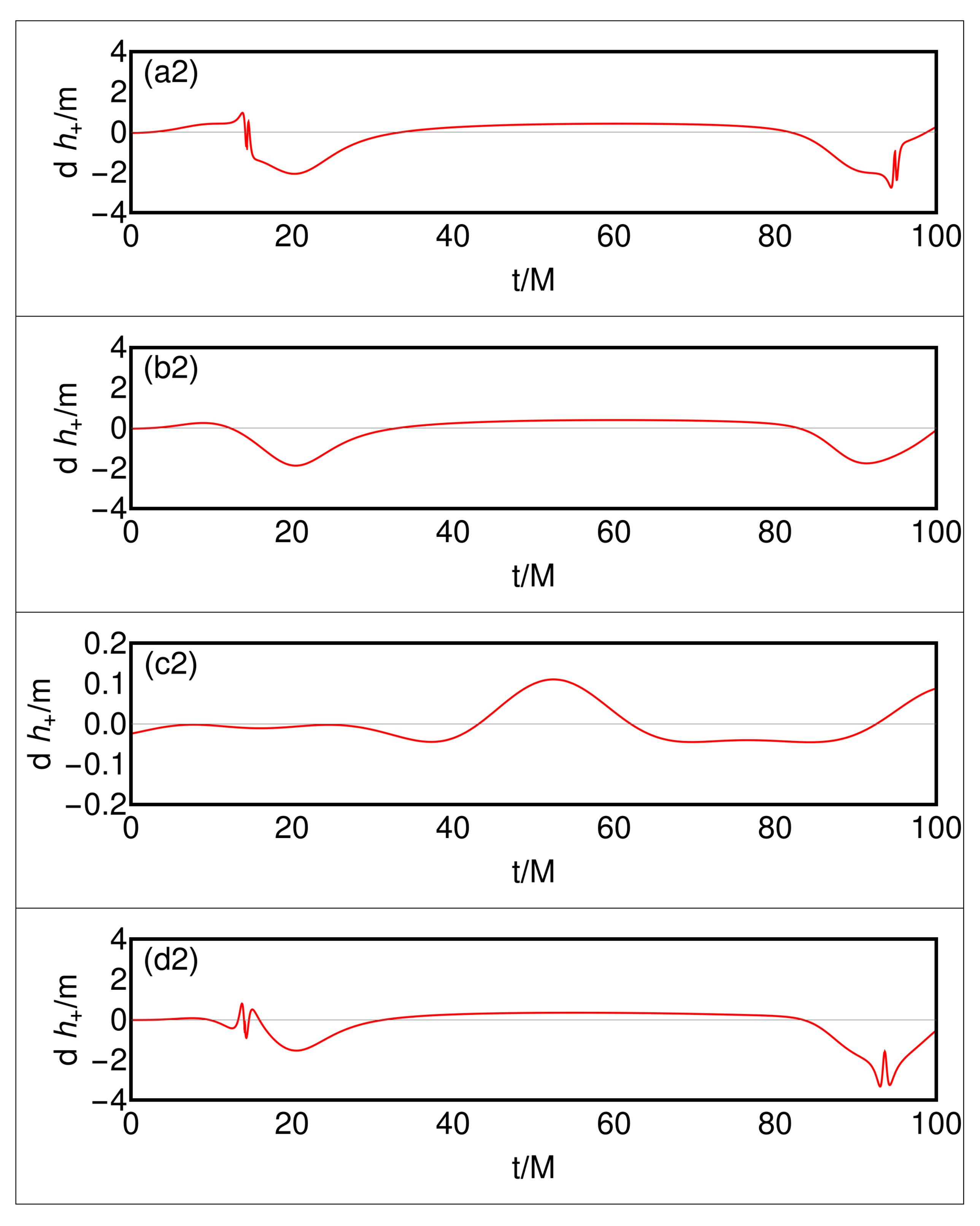}
			\end{center}
			\vspace{-5mm}
			\caption{Amplitudes of the plus polarization for the GWs generated by the orbits in Fig.~\ref{orbits-123}.}
			\label{gw-waveform-90-abcd}
		\end{figure}
		
	\end{widetext}

	In principle, the information of the {waveforms} should show the properties of the corresponding orbits. For the orbits {obtained} in Fig.~\ref{orbits-123}, we expect that we could observe {novel GWs}, especially for the orbits (c) and (d). {Figure}~\ref{gw-waveform-90-abcd} shows the time-domain GWs generated by the orbits in Fig.~\ref{orbits-123} viewed at an inclination angle $90^\circ$, where the results in right panel (a2, b2, c2, d2) correspond to the zooms of the GWs in left panel (a1, b1, c1, d1). Checking the corresponding GWs in Fig.~\ref{gw-waveform-90-abcd}, we find that there are quasi gravitational radiation pulses generated by the orbits (a) and (d), and it is absent for the orbits (b) and (c).
	
\begin{figure*}[htbp]
		\begin{center}
			\includegraphics[width=1\linewidth]{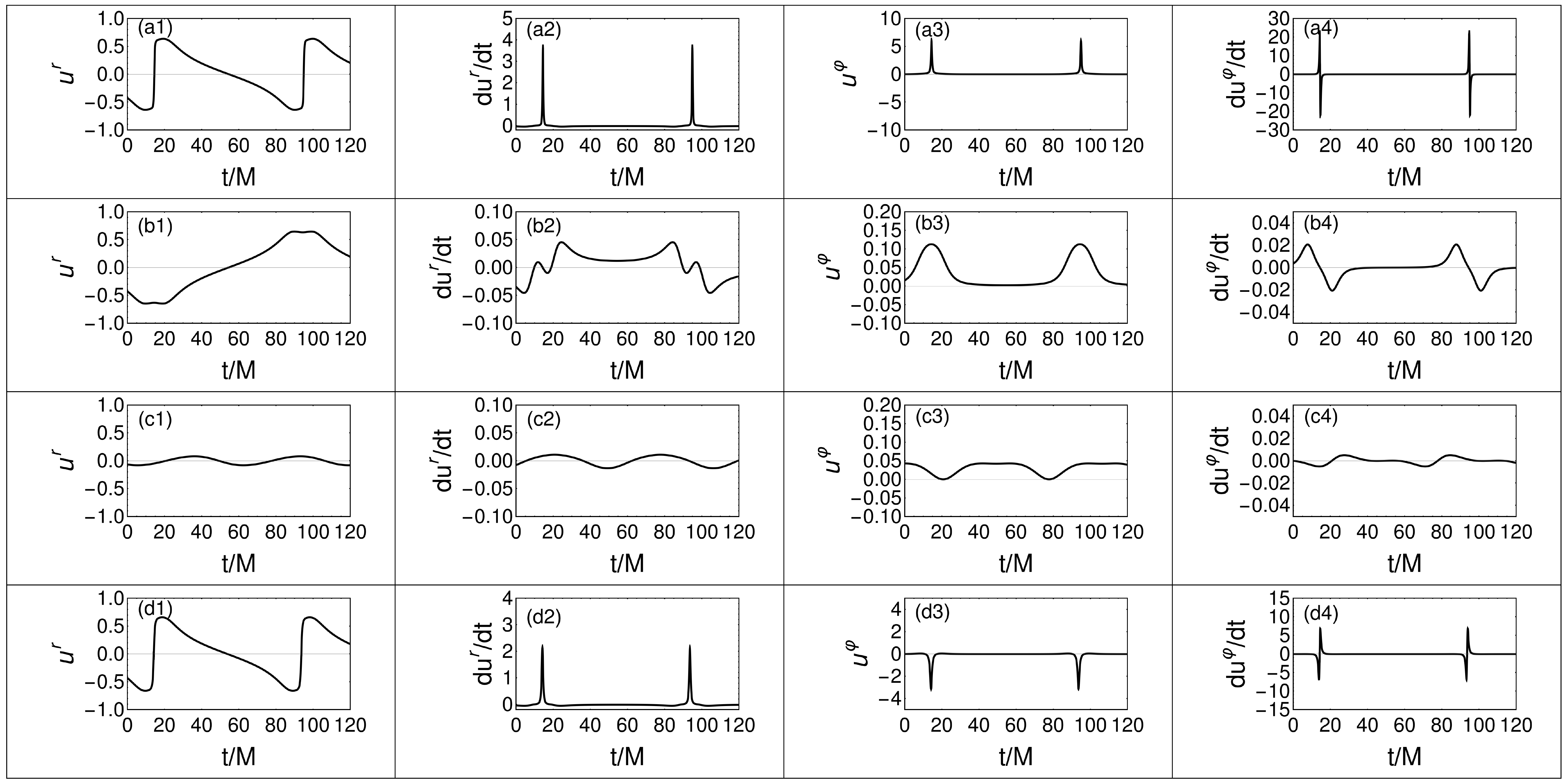}
		\end{center}
		\vspace{-5mm}
		\caption{Velocity $(u^r, u^\varphi)$ and acceleration $(du^r/d \tau, du^\varphi/d \tau)$ of the orbits in Fig.~\ref{orbits-123}. The results in subfigures (i1-i4) correspond to the orbit (i), where $i=(a, b, c, d)$.}
		\label{vrvphi}
	\end{figure*}

	In view of the orbits in Fig.~\ref{orbits-123} and the GWs in Fig.~\ref{gw-waveform-90-abcd}, the dependence of the corresponding GWs and orbits does not seem to be the same as our expectation. By analyzing the corresponding velocities and accelerations at different positions along the orbits of the test particle, we find that although there are peaks in {the} orbits (c) and (d), only the orbit (d) can generate gravitational radiation pulses. In addition, the orbit (a) can also generate gravitational radiation pulses. To get the relations between the GWs and orbits, we go back to the definition of the waveform \eqref{eq-gws}. Following Eq. \eqref{eq-gws} we know that the amplitudes of the GWs are proportional to the second-order time derivative of the quadrupole, which means the GWs are dependent on the orbit location, {velocity, and acceleration} of the test particle. Note that the orbital eccentricities of the orbits (a) and (d) are close to but not equal to $1$, the corresponding velocity $(u^r, u^\varphi)$ will change greatly in a very short time when the test particle is passing through the periastron $r_p$. This change is the main cause of gravitational radiation pulses. {Figure}~\ref{vrvphi} shows the velocity $(u^r, u^\varphi)$ and acceleration $(du^r/d \tau, du^\varphi/d \tau)$ of each orbit given in Fig.~\ref{orbits-123}. These behaviors are not possessed by the supermassive black holes. Combined with the future space-borne gravitational wave detection plan, it could be possible to verify whether {there are supermassive boson stars in galactic centers}.
	
	Next, we discuss the detectability of the GW signals we obtained. The corresponding mass of the rotating boson {star is}
	\begin{equation}
	M\approx 1.30 \frac{m_p^2}{\mu}.
	\end{equation}
Here, we should note that, the prediction of Ref. \cite{Amaro-Seoane:2010pks} showed that the mass parameter of the scalar field is of about $\mu \in [300\,\text{eV},\,2\times 10^4\,\text{eV}]$. This prediction is obtained in terms of the astronomical observations of Sgr $A^*$, NGC 4258, and Collisional dark matter. However, it is shown that the total number of galaxies in the univesrse up to $z=8$ is almost $10^{12}$ \cite{Christopher2016}. Therefore, faced with such a large number of galaxies, we cannot strictly rule out the absence of supermassive boson stars in the galactic centers. If we set the mass parameter of the scalar field to be $\mu \approx 10^{-16}\text{eV}$, then the mass of the rotating boson star can be $M=10^6M_\odot$.

\begin{figure}[htbp]
		\begin{center}
			\includegraphics[width=0.8\linewidth]{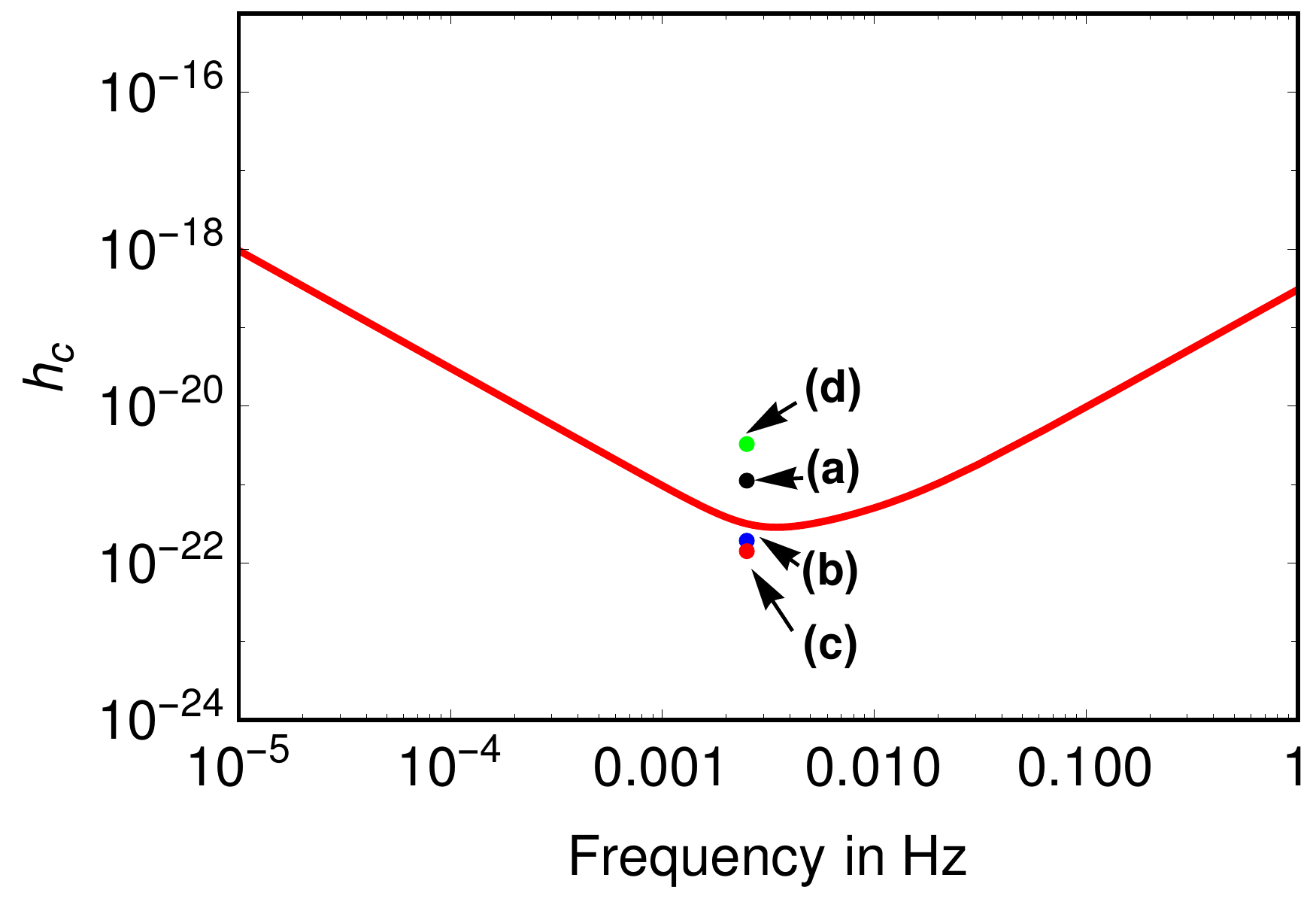} \hspace{5mm}
		\end{center}
		\vspace{-5mm}
		\caption{Characteristic amplitude of the GW signals given in Fig.~\ref{gw-waveform-90-abcd} and LISA's sensitivity window~\cite{Moore2015}, where the points are determined by the maximal amplitudes and frequencies of the whole GW signals, (a-d) indicates the corresponding orbits of the GW signals.}
		\label{lisa-sensitivity-emri-signals}
	\end{figure}

	We set the mass of the small star to be $m=10 M_\odot$. The distance from the detectors to the source is set up to $1 \text{Gpc}$. With such assumptions, we could estimate the maximal amplitudes and frequencies of the GWs. We find that the frequencies of the pulse part could be in the magnitude of $10^{-1}$Hz, while the frequencies of the whole GW signals are still in the milli-Hz. {By naively comparing} the LISA's sensitivity window and the GW signals in Fig.~\ref{lisa-sensitivity-emri-signals}, we find that these novel GW signals could be detected by the space-borne gravitational-wave detectors.
	
	\section{Conclusions}
	
	In this work, we obtain the GWs with the quadrupole approximation for an EMRI
	system with a central supermassive rotating boson star. We consider four types
	of special orbits (Fig.~\ref{orbits-123}) and derive the corresponding GWs. Due
	{the horizonless} property of the rotating boson star, the stellar-mass compact
	object could pass through the center of the rotating boson star and the orbital
	eccentricity could be $e=1$. The
	distribution of the scalar field in the whole space also allows the
	stellar-mass compact object to be at rest for some special orbits (the orbits
	(c) and (d) in Fig.~\ref{orbits-123}). We calculate the corresponding GWs from
	the orbits in Fig.~\ref{orbits-123} {and find} that {there are} gravitational
	radiation pulses when the stellar-mass compact object moves along the orbits
	(a) and (d).  Although the orbits (b) and (c) have peaks, there are no
	gravitational radiation pulses generated by them. These novel GWs are quite
	different from the EMRI system with a central supermassive Kerr black hole, and
	can therefore be envisaged as a way to discover supermassive rotating boson
	stars. Assuming for the boson star and stellar compact object masses
	$M=10^6M_\odot$ and $m \approx 10 M_\odot$, approximately, and assuming a
	distance of $d=1{\text{Gpc}}$ between the source and detector we showed that such system could be detected by the future space-borne
	gravitational-wave detectors.
	
	Although our numerical method is not accurate enough compared with the
	perturbation method, the research provides a stepping stone for the study of supermassive boson stars in galaxies with the help of the future space
	gravitational-wave detection.

	
	\section{Acknowledgments}
	This work was supported in part by the National Key Research and Development Program of China (Grant No. 2020YFC2201503), the National Natural Science Foundation of China (Grants No. 11875151, No. 11705070, No. 12105126, and No. 12075103), the 111 Project under (Grant No. B20063), the Fundamental Research Funds for the Central Universities (Grants No. lzujbky-2021-pd08 and lzujbky-2019-ct06), and ``Lanzhou City's scientific research funding subsidy to Lanzhou University''. 
	PAS acknowledges support from the Ram{\'o}n y Cajal Programme of the Ministry
	of Economy, Industry and Competitiveness of Spain, as well as the financial
	support of Programa Estatal de Generación de Conocimiento (ref.
	PGC2018-096663-B-C43) (MCIU/FEDER). This
	work was supported by the National Key R\&D Program of China (2016YFA0400702)
	and the National Science Foundation of China (11721303).

\end{document}